# Stabilizing polar phases in binary metal oxides by hole doping

Tengfei Cao[1,2]*, Guodong Ren[3], Ding-Fu Shao[4], Evgeny Y. Tsymbal[1]*, and Rohan Mishra[2,3]*

[1]Department of Physics and Astronomy & Nebraska Center for Materials and Nanoscience, University of Nebraska, Lincoln, Nebraska 68588-0299, USA

[2]Department of Mechanical Engineering & Materials Science, Washington University in St. Louis, St. Louis, MO 63130, USA

[3]Institute of Materials Science & Engineering, Washington University in St. Louis, St. Louis, MO 63130, USA

[4]Key Laboratory of Materials Physics, Institute of Solid State Physics, HFIPS, Chinese Academy of Sciences, Hefei 230031, China

**Abstract**

The recent observation of ferroelectricity in the metastable phases of binary metal oxides, such as $HfO_2$, $ZrO_2$, $Hf_{0.5}Zr_{0.5}O_2$, and $Ga_2O_3$, has garnered a lot of attention. These metastable ferroelectric phases are typically stabilized through epitaxial growth, alloying, or defect engineering. Here, we propose hole doping plays a key role in stabilizing the polar phases in binary metal oxides. Using first-principles density-functional-theory calculations, we show that holes in these oxides mainly occupy one of the two oxygen sublattices. This hole localization, which is more pronounced in the polar phase than in the nonpolar phase, lowers the electrostatic energy of the system, and makes the polar phase more stable at sufficiently large concentrations. We demonstrate that this electrostatic mechanism is responsible for stabilization of the ferroelectric phase of $HfO_2$ aliovalently doped with elements that introduce holes to the system, such as La and N. Finally, we show that the spontaneous polarization in $HfO_2$ is robust to hole doping, and a large polarization persists even under a high concentration of holes.

**Introduction**

Binary metal oxides (BMOs) with wide bandgaps and large dielectric constants, such as $SiO_2$ and $HfO_2$, are used as gate dielectrics in transistors[1-2]. Recently, ferroelectricity has been reported in various BMOs such as $HfO_2$, $ZrO_2$ and $Ga_2O_3$[3-9]. Their ferroelectric phases are, however, metastable compared to a nonpolar phase that is most stable under ambient conditions[10-14]. There are several strategies to stabilize the metastable ferroelectric phases: doping[7, 10-11, 15-23], alloying[18, 20, 24-25], defect engineering[26-32], and strain engineering through epitaxy with substrates[33-42], with the doping being the most common approach to be used experimentally. For example, the introduction of Al into $Ga_2O_3$ stabilizes its ferroelectric $ε$-phase[10]. For $HfO_2$, there are even more experimental explorations of polarization and stability of its ferroelectric phase. For example, Toriumi *et al*. systematically investigated the effect of doping, at both the cation and anion sites, on the polarization of HfO2. They concluded that hole dopants contributed to an increase in the volume fraction of the ferroelectric phase over its non-polar phases[43]. Hwang *et al* also observed an increase proportion of the ferroelectric phase in $Hf_{0.5}Zr_{0.5}O_2$ thin films in contact with a layer of $HfO_xN_y$, which they attributed to stabilization due to N-doping[44-45], and the ferroelectric endurance of $HfO_2$-$ZrO_2$ could be enhanced by La doping[46]. Song *et al*. have also recently reported a higher remnant polarization in $Hf_{0.5}Zr_{0.5}O_2$ films doped with La[47]. Apart from all above work, various other experimental studies also show that Si, Zr, La, Sc, Y, and N facilitate the formation of orthorhombic ferroelectric phase of $HfO_2$[18, 20, 25]. Oxygen vacancies, especially in ordered form, have also been proposed to stabilize the orthorhombic phase of $HfO_2$[48]. Finally, these oxides are regularly grown as thin films on carefully selected substrates that can impose adequate strain for the stabilization of the ferroelectric phase[14, 33].

All the above strategies to stabilize metastable ferroelectric phases involve the transfer of charges. For instance, dopants donate or accept electrons, and oxygen vacancies typically act as electron donors. If there is a mismatch in the work function of the substrate and the film, or if the substrate has a different polarity, it can result in the transfer of charges to the film[49-50]. There are a handful of theoretical reports that have explored the presence of charges and their impact on the crystal structure and electronic properties[51-54]. Especially for hole doping, the large electronegativity difference between the metal and oxygen ions in BMOs can localize holes spatially in all oxygen sublattices. Even oxygen vacancy may locally neutralize hole states in a BMO matrix. Extra holes spreading in remaining oxygen lattices could impact structure and electronic properties of metal oxides. For example, McKenna *et al.*[52], have shown, using electronic structure calculations, that extra holes in $HfO_2$ or $ZrO_2$ could arrange as two-dimensional (2D) conductive sheets. Muñoz Ramo *et al.*[51], have predicted that holes exist as polaronic states in monoclinic $HfO_2$ by distorting the lattice locally. Free carriers, either in the form of holes or electrons, have been observed to stabilize metastable phases. For example, charge carriers can screen the polarization and suppress the associated polar distortions in prototypical displacive perovskite ferroelectrics such as $BaTiO_3$, and induce a phase transition from the ferroelectric to the paraelectric phase[55]. Conversely, a recent work on hybrid-improper ferroelectrics shows that free carriers can strengthen polar distortions and facilitate the formation of ferroelectric phases[56]. Various mechanisms have been proposed to explain the effect of doping on the polarization in traditional ferroelectrics, including, meta-screening and second-order Jahn-Teller effects. However, compared to traditional perovskite ferroelectrics, the effect of charge doping, especially holes, on the phase stability of newly discovered polar phases in binary metal oxides remain poorly understood. Especially for polar metal oxides, extra charges could not only arrange into specific patterns that partially break metal-oxide bonds, but also highly couple with polar displacements and greatly modulate their structures and phase stabilities. Therefore, whether electron or hole doping can stabilize a metastable polar phase and permit ferroelectric switching remains an open question.

In this work, we analyze the effect of charge doping on the stability of the metastable polar phase in BMOs and the evolution of their electronic properties. We find that holes can stabilize a metastable polar phase in BMOs having two different oxygen sublattices and a nonpolar ground state such as $HfO_2$, $Hf_{0.5}Zr_{0.5}O_2$, $ZrO_2$, and $Ga_2O_3$. This hole-doping-induced ferroelectric stability arises from the localization of holes, which reduces the long-range electrostatic interactions. Specifically, holes are found to preferentially localize at triply coordinated oxygen sites (O3) favoring them over oxygen sites having four nearest cation neighbors (O4). The holes can even arrange into 2D sheets within the three-dimensional structure. This spatial arrangement into 2D sheets formed by the O3 ions reduces the electrostatic energy of the doped polar phase with respect to the nonpolar phase. We also find that the spontaneous polarization persists in the presence of the localized holes. It results in a small decrease in the height of the ferroelectric double-well barrier, which is often associated with the switching field or the ferroelectric-paraelectric transition temperature. In the case of $HfO_2$, we find that a hole concentration of $3.86 \times 10^{21}$ cm$^{-3}$ can stabilize the polar orthorhombic phase. Such hole concentrations can come either through doping or ionic gating, especially during the initial growth of $HfO_2$. Our work suggests that hole doping, either intentionally through ionic gating, or unintentionally through doping and alloying plays an important role in the stabilization of the metastable polar phases.

**Methods**

We carried out density-functional-theory (DFT) calculations using the Vienna *Ab-initio* Simulation Package (VASP)[57]. The energy cutoff for the plane waves was set at 500 eV. The threshold for energy convergence for the self-consistent loops was set at $10^{-6}$ eV. For structural optimization, the convergence of forces was set to 0.001 eVÅ$^{-1}$. We used projector augmented-wave (PAW) potentials[58] and the generalized gradient approximation (GGA) within the Perdew-Burke-Ernzerhof (PBE)[59] parameterization to describe the electron-ion and electronic exchange-correlation interactions, respectively. The Brillouin zone was sampled using the Monkhorst-Pack method with the smallest allowed spacing between *k*-points set to 0.1 Å$^{-1}$[60]. The optimized in-plane lattice parameters of the ferroelectric and the nonpolar phases of HfO$_2$, Hf$_{0.5}$Zr$_{0.5}$O$_2$, ZrO$_2$, Ga$_2$O$_3$, and their space groups are given in Table S1. To simulate charge doping, we changed the total number of electrons in the BMO unit-cells and optimized the charged structures under a homogeneous charge background. Cation or anion alloying effects in HfO$_2$ were simulated using the virtual crystal approximation method[61-62], where La was chosen as a cation and N, P, and Sb as anions.

The electrostatic interactions between doped charges in HfO$_2$, Hf$_{1-x}$La$_x$O$_2$, and Ga$_2$O$_3$ were simulated using a core-shell model[63]. We used the atomic structures of HfO$_2$, Hf$_{1-x}$La$_x$O$_2$, and Ga$_2$O$_3$ that were fully optimized with DFT and imposed periodic boundary conditions. In the core-shell, each atomic site encompassed an ionic charge $Q_i$ fixed to the equilibrium position and an electronic charge $q_i$ bound to its ionic site. The ionic charges were represented by point charges. The electronic charges $q_i$ were adjusted to simulate hole distributions on oxygen sublattices and broadened by a Gaussian distribution of width $\sigma_i$. All the parameters were fitted to reproduce the dielectric constants and energetics of polar and nonpolar phases of HfO$_2$ and Ga$_2$O$_3$ in the charge-neutral optimized states and are given in Table S1 and Table S6. The electrostatic energy was calculated using the Ewald method[63]:

$$U_{Ewald} = \frac{1}{2}\sum_n \sum_i^{N+n} \sum_{j \neq i}^{N+n} \frac{q_i q_j}{|r_i - r_j - R_n|} erfc\left[\frac{r_i - r_j - R_n}{\sqrt{2}\sigma_{ew}}\right]$$
$$+ \frac{2\pi}{V}\sum_{G \neq 0} \frac{\exp\left(-\frac{\sigma_{ew}^2 G^2}{2}\right)}{G^2} |S(G)|^2 - \frac{1}{\sqrt{2\pi}\sigma_{ew}} \sum_i^{N+n} q_i^2$$
$$+ \sum_i^n \frac{q_i Q_i}{\Delta s_i} erf\left(\frac{\Delta s_i}{\sqrt{2}\sigma_{ew}}\right)$$
$$- \sum_n \sum_i^n \sum_j^N \frac{q_i Q_j}{|s_i - r_j - R_n|} erfc\left[\frac{s_i - r_j - R_n}{\sqrt{2}\sigma_i}\right]$$
$$- \sum_n \sum_i^n \sum_{j \neq i}^n \frac{q_i q_j}{|s_i - r_j - R_n|} erfc\left(\frac{s_i - r_j - R_n}{\sqrt{2(\sigma_i^2 - \sigma_j^2)}}\right)$$

(Eq. 1)

where *V* is the volume of supercell, n and $R_n$ are indexes and corresponding lattice vector of the direct space of supercells, $G$ is a vector in the reciprocal space, $r_i$ is a vector representing the position of charge $q_i$, $S(G)$ is the structure factor, $\sigma_{ew}$ is the parameter controlling convergence of the Ewald sums, and $\Delta s_i$ is the distance between an ion and the respective electronic charge. The first sum excludes the interactions between ions and electronic charges belonging to the same site, and the fourth sum is used to eliminate these same interactions from the sum in the reciprocal space. The electrostatic energy variation due to the extra holes is calculated using $E_s = E_s(h) - E_s(n)$, where $E_s(h)$ and $E_s(n)$ are the electrostatic energies of the hole-doped and charge-neutral states, respectively.

To obtain the hole charges at a given site for different levels of doping, we used the difference in the Bader charges[10] between the charge-doped and charge-neutral systems. We calculated the deformation energy from the difference in the DFT total energy between the deformed structure and the most stable structure in the charge-neutral state. The deformed structure was obtained by a full optimization of the lattice and the atomic positions under different charge doping levels. We then kept the optimized structure and removed the extra charge in the matrix to calculate its energy with respect to the most stable structure in the charge-neutral state. All energies are calculated with reference to the energy of the most stable monoclinic structure. We obtain the ferroelectric double-wells in Fig. 5a based on group-theoretical analysis. For the ferroelectric phase of $HfO_2$ with space group $Pca2_1$, we obtain its centrosymmetric structure with $Pcca$ space group, which is a supergroup of $Pca2_1$, with minimal atomic distortions. Using the Pseudo program of the Bilbao Crystallographic Server[64], the $Pca2_1$ sub-group can be obtained from the $Pcca$ supergroup by freezing the $\Gamma_3^-$ polar phonon mode. We obtain the double-well in Fig. 5a by freezing different amplitude of this $\Gamma_3^-$ mode. We calculated the polarization along this distortion mode using the Berry phase method[65] and obtained a value of 51.4 μC/cm$^2$, which is consistent with previous theoretical results[66].

**Results**

The structures of $HfO_2$ and $Ga_2O_3$, both in their metastable orthorhombic polar phase (o-$HfO_2$ and ε-$Ga_2O_3$) and their ground-state monoclinic phase (m-$HfO_2$ and β-$Ga_2O_3$), are shown in Fig. 1(a). The polar and most stable nonpolar crystal structures of $Hf_{0.5}Zr_{0.5}O_2$ and $ZrO_2$ are the same as those of $HfO_2$. All these structures have two symmetry inequivalent oxygen sub-lattices. The oxygen sub-lattice with four-fold and three-fold coordination are shown in red and yellow colors, respectively. The lattice parameters, space group, and formation enthalpy of nonpolar and polar phases are given in Tables S1 and S6, respectively. Those nonequivalent atomic sites for each polar and nonpolar phase are shown in Table S3 and Table S8. For comparison, lattice parameters and atom coordinates of polar and nonpolar phases from available experimental results are also exhibited in supplementary materials (Tables S2, S4, S7 and S7). The energy difference between $HfO_2$, $Hf_{0.5}Zr_{0.5}O_2$, $ZrO_2$, and $Ga_2O_3$ in their polar phase and their ground state nonpolar phase is plotted in Fig 1(b), as a function of carrier doping ranging from –0.24 to 0.36 h/f.u., which corresponds to doping concentration from –7.14×10$^{21}$ to 0.11×10$^{23}$ h/cm$^{-3}$. In the charge-neutral optimized state, the energy of the polar phase of o-$HfO_2$ is 0.082 eV/f.u. higher than nonpolar m-$HfO_2$. The energy of polar o-$Hf_{0.5}Zr_{0.5}O_2$ is 0.081 eV/f.u. higher than that of m-$Hf_{0.5}Zr_{0.5}O_2$, and the energy of the polar structure of ε-$Ga_2O_3$ is 0.15 eV/f.u. higher than that of nonpolar β-$Ga_2O_3$. These results are consistent with the values reported in the literature[12, 14]. They are also consistent with the experimental observations, which show that large-area pure polar phases of $HfO_2$, $Hf_{0.5}Zr_{0.5}O_2$, and $Ga_2O_3$ can hardly be achieved, and proper substrates, doping, and specific growing conditions are prerequisites to stabilize the polar phase[8, 18, 23, 67].

We find that charge doping has a large effect on the relative stability of the polar phase with respect to the nonpolar phase for all the four BMOs, as shown in Fig. 1(b). While electron doping makes the nonpolar phase more stable, hole doping stabilizes the polar phase of all the four BMOs. To understand the mechanism for stabilization of the polar phase with hole doping, we first analyze the electronic structure and determine whether there is any localization of the added holes in both phases. Here, we use $HfO_2$ and $Ga_2O_3$ as representative examples as $Hf_{0.5}Zr_{0.5}O_2$ and $ZrO_2$ show similar behavior. We focus on the O 2$p$ states as they make up the valence band. In Figs. 2(a,b), we show the O-2$p$ density of states (DOS) in o-$HfO_2$ and m-$HfO_2$, both in their neutral state and at a hole-doping level of 0.2 h/f.u., where polar o-$HfO_2$ becomes more stable. As mentioned above, there are two types of oxygen sublattices in these BMOs, comprising the triply coordinated O3 sites and the four-fold coordinated O4 sites. In both o-$HfO_2$ and m-$HfO_2$, holes preferably occupy the 2$p$ states of the O3 sublattice, as can be seen in Figs. 2 (a, b). A comparison of the O 2$p$ DOS for o-$HfO_2$ and m-$HfO_2$ shows that the added holes are more localized in energy in the polar o-$HfO_2$. We observe by the sharp edge in o-$HfO_2$ at the Fermi energy (Fig. 2(a)), whereas in nonpolar m-$HfO_2$, the hole states span a larger energy range (Fig. 2(b)). Therefore, at the same level of hole doping, the Fermi energy shift with respect to the valence band edge of the charge-neutral HfO$_2$ is larger in $m$-HfO$_2$ than in $o$-HfO$_2$. We observe a similar tendency in Ga$_2$O$_3$. Figs. 2(c,d) show the O-2$p$ DOS for $\varepsilon$-Ga$_2$O$_3$ and $\beta$-Ga$_2$O$_3$ in the charge-neutral state and at a hole-doping level of 0.2 h/f.u., where polar $\varepsilon$-Ga$_2$O$_3$ becomes more stable. It is seen that the holes preferentially occupy the O3 sublattice, and the 2$p$ states of oxygen in polar $\varepsilon$-Ga$_2$O$_3$ are more localized in energy than in $\beta$-Ga$_2$O$_3$.

The differences in the degree of hole localization in polar and nonpolar phases of HfO$_2$ and Ga$_2$O$_3$ are evident from the iso-surface plots of the hole density, shown in Figs. 2(e,f). In both the compounds, the hole density is more spatially localized in the polar phase than in the nonpolar. Specifically, in the case of HfO$_2$, the added holes arrange themselves in parallel, quasi-2D hole sheets formed by the O3 sublattice in both $m$-HfO$_2$ and $o$-HfO$_2$ (Fig. 2(e)). However, while the hole density is continuous in $m$-HfO$_2$ along the $z$-axis, it is discontinuous in $o$-HfO$_2$. Also, along the $y$-axis, the spatial separation between the hole sheets in $m$-HfO$_2$ is 2.26 Å which is shorter than 2.35 Å in $o$-HfO$_2$. The more delocalized distribution of the holes in $m$-HfO$_2$ manifests itself in a larger dispersion of the valence band edge, both in energy and in the reciprocal space along the Γ-M and Γ-A directions, as shown in Fig. S1. Similar tendency is observed in Ga$_2$O$_3$, where the hole density is localized in a large hollow site of polar $\varepsilon$-Ga$_2$O$_3$, whereas it is more uniformly distributed in nonpolar $\beta$-Ga$_2$O$_3$ (Fig. 2(f)).

These different hole distributions in polar and nonpolar phases of HfO$_2$ and Ga$_2$O$_3$ affect their relative stability and eventually stabilize the polar phase at sufficiently high hole doping. We analyze the structural deformation and the electrostatic energy variation to validate this argument. We first evaluate the effect of structural distortions associated with the additional holes on the relative stability of the polar and nonpolar phases using HfO$_2$ and Ga$_2$O$_3$ as examples. To do this, we fully optimize the structure (both lattice and ions) under each doping level, then we remove the extra charges and calculate the energy of the optimized structure (obtained under charge doping) in the charge-neutral state. These structures have been deformed by extra charges in their matrices with respect to the fully optimized charge-neutral structure, so we use the deformation energy to show the effect of structural distortions on the relative stability of the polar and nonpolar phases. The results shown in Figs. 3(a,b) clearly indicate that, for both polar and non-polar phases, their energies increase with the structure deformations. However, the relative stability of the polar and nonpolar phase does not change, and the nonpolar monoclinic structure of both HfO$_2$ and Ga$_2$O$_3$ remains the lower energy state.

It appears that it is the electrostatic energy difference between the polar and the nonpolar phases upon hole doping that stabilizes the polar phase in HfO$_2$ and Ga$_2$O$_3$. We calculate the electrostatic energy using the core-shell model as described in the Methods section. In the calculations, we assume, for simplicity, that the extra holes solely occupy the O3 sublattice, reflecting qualitatively our DFT results (Fig. 2). The results are shown in Figs. 3(c) and 3(d), for HfO$_2$ and Ga$_2$O$_3$, respectively. It is seen that the electrostatic energy gradually increases with the increasing number of holes for both the phases. However, this increase is greater for the nonpolar phase than for the polar phase, eventually making the latter more stable. This behavior of the electrostatic energy as a function of hole doping is consistent with the hole density being more localized on the O3 sites in the polar phases compared to the nonpolar (Fig. 2). Thus, our electrostatic model qualitatively reproduces the tendency derived from the DFT total energy calculations (Fig. 1(b)).

Experimentally, the extra holes could come from ionic gating or from the substrate. They could also come from ionic dopants and aliovalent alloying. Indeed, experimentally, additions of La, Y, Sc, or N to HfO$_2$ have been reported to stabilize its ferroelectric phase ($o$-HfO$_2$)[10, 68-69]. All these elements introduce holes into the system. To analyze whether the mechanism discovered above also applies to such $p$-type dopants and alloys, we consider the effect of La doping on the relative phase stability of $o$- and $m$-HfO$_2$.

We observe that on increasing the La concentration, the energy difference between $o$- and $m$-$HfO_2$ phases gradually decreases, until at $x_{La} > 0.35$ in $Hf_{1-x}La_xO_2$, ferroelectric $o$-$HfO_2$ becomes more stable than $m$-$HfO_2$, as shown in Fig. 4(a). The hole distribution at $x = 0.35$ is shown in Fig. 4(b). As seen from the DOS plots in this figure, the holes primarily occupy the O3 sublattice and are more localized in $o$-$HfO_2$ than in $m$-$HfO_2$. This trend remains for the entire concentration of La considered here, as seen from the hole occupancy at the O3 and O4 sites for the two phases as a function of $x$ (0.00–0.40) in Fig. 4(c). Interestingly, the holes from La ions are mainly localized at the O3 sites, even though the La atoms form chemical bonds with both O3 and O4. Using the hole distribution in Fig. 4(c), we calculate the electrostatic energy as a function of $x$ for the two phases of $Hf_{1-x}La_xO_2$ using the core-shell model. The results shown in Fig. 4(d) indicate that with increasing $x$, the electrostatic energy increases more drastically for $m$-$Hf_{1-x}La_xO_2$ than for $o$-$Hf_{1-x}La_xO_2$. At $x = 0.35$, the electrostatic energy difference becomes ≈ 0.08 eV/f.u., which is sufficient to stabilize the orthorhombic phase according to our total energy calculation shown in Fig. 4(a). These results demonstrate that for dopants introducing holes to $HfO_2$, it is less increase of electrostatic energy due to the hole localization that stabilizes the polar phase. Apart from the cation doping, we have also analyzed substitution of O by N, which introduces holes to the system, on the relative stability of the two phases in $HfO_2$. The results are similar to those observed for La alloying, as shown in Fig. S2. Together, these results demonstrate that as long as holes are localized on O3 sublattice, they can facilitate the stabilization of the polar phase.

Finally, we analyze the effect of hole doping on the switching barrier and polarization of $o$-$HfO_2$, using $o$-$Hf_{1-x}La_xO_2$ as an example. We calculate the change in the ferroelectric double-well barrier following transition from the $o$-phase with the $Pca2_1$ space group and polarization pointing one way, through a centrosymmetric, nonpolar structure with the $Pcca$ space group, to the $Pca2_1$ phase with the polarization reversed with respect to the initial state, as shown in Fig. 5 (c). This barrier height often correlates well with the coercive field required for ferroelectric switching and the ferroelectric to paraelectric transition temperature. As seen from Fig. 5 (a), the height of the barrier between the polar and nonpolar phases reduces upon hole doping with La addition. For pure $o$-$HfO_2$, the switching barrier is 0.29 eV/f.u. It reduces to 0.18 eV/f.u. for $o$-$Hf_{0.9}La_{0.1}O_2$, 0.13 eV/f.u. for $o$-$Hf_{0.8}La_{0.2}O_2$, and 0.09 eV/f.u. for $o$-$Hf_{0.7}La_{0.3}O_2$. We have also estimated the polarization under each doping level. In the calculations, $o$-$Hf_{1-x}La_xO_2$ with any finite concentration of La exhibits metallic-like properties (Fig. 4b), so we cannot calculate ferroelectric polarization using the Berry-phase approach[65]. Instead of polarization, we obtain the magnitude of polar displacement at each doping level, which are known to correlate with the polarization[66]. We find that the polar displacement in $o$-$Hf_{1-x}La_xO_2$ is robust to La addition. Even under very high concentration of La, $x = 0.3$, the polar displacements remain as large as 0.46 Å, compared to 0.54 Å in pure $o$-$HfO_2$. Based on these displacements, we estimate the ferroelectric polarization of $o$-$HfO_2$. We find that, for pure $o$-$HfO_2$, the polar displacement is 0.54 Å, and its ferroelectric polarization calculated using the Berry-phase approach is 51.4 μC/cm². Assuming a linear relationship between the polar displacements and polarization, we expect the polarization at $x = 0.3$ to be as large as 41.2 μC/cm². Therefore, under hole doping induced by La dopants, not only the ferroelectric phase of $HfO_2$ can be stabilized, but its spontaneous ferroelectric polarization is also preserved.

$HfO_2$ is unique because its ferroelectricity is retained even in ultrathin films[26, 70]. Various cation and anion dopants, such as La, Sc, Y, and N, have been used to stabilize the ferroelectric phase of $HfO_2$, and all these elements donate holes into the $HfO_2$ matrix. Moreover, ferroelectric films of $Hf_{0.5}Zr_{0.5}O_2$ have been synthesized on silicon[26] or $La_{2/3}Sr_{1/3}MnO_3$ substrates[33]. Interestingly for $La_{2/3}Sr_{1/3}MnO_3$, it is observed that the ferroelectric phase of $Hf_{0.5}Zr_{0.5}O_2$ could only be synthesized on the $MnO_2$ terminated surface. We calculated the work function of different surface to analyze possible hole transfer from substrate to epitaxial $Hf_{0.5}Zr_{0.5}O_2$ film. We find that the work function of $Hf_{0.5}Zr_{0.5}O_2$ is −3.45 eV. The surface work function of silicon is in the range from −4.60 eV to −4.85 eV. The work function of the $MnO_2$-terminated $La_{2/3}Sr_{1/3}MnO_3$ surface is −5.94 eV. The lower work function of the substrates with respect to $Hf_{0.5}Zr_{0.5}O_2$ suggests that holes can transfer across the interface to $HfO_2$ and stabilize its metastable ferroelectric $o$-phase. These experimental results indicate the stabilizing role of holes in ferroelectric phase formation in $Hf_{0.5}Zr_{0.5}O_2$. Thus, our findings here provide new insights into these experimental findings as well as give hints to experimentalists for exploring more efficient approaches to stabilize the polar phases of $HfO_2$ and other binary compounds such as $Ga_2O_3$, $HfZrO_2$, and $ZrO_2$.

**Summary**

In this work, we have identified a mechanism through which holes can stabilize polar phases in BMOs using a combination of DFT calculations and core-shell models. The injected holes preferentially occupy the O3 sublattice having triply coordinated oxygen atoms. This hole localization, which is more pronounced in the polar phase than in the nonpolar phase, in turn lowers the electrostatic energy of the system, and makes the polar phase more stable at sufficiently large hole concentrations. We find that this behavior is also observed for aliovalent alloying with elements, such as La and N, that introduce holes to the system. Furthermore, we find that the switching barrier of $o$-$Hf_{1-x}La_xO_2$ is reduced with the increasing number of holes, but its spontaneous polarization persists even at such high hole concentrations. Our findings contribute to the understanding of the ferroelectric phase formation in BMOs and pave the way to stabilize ferroelectric phases of BMOs.

**Figures**

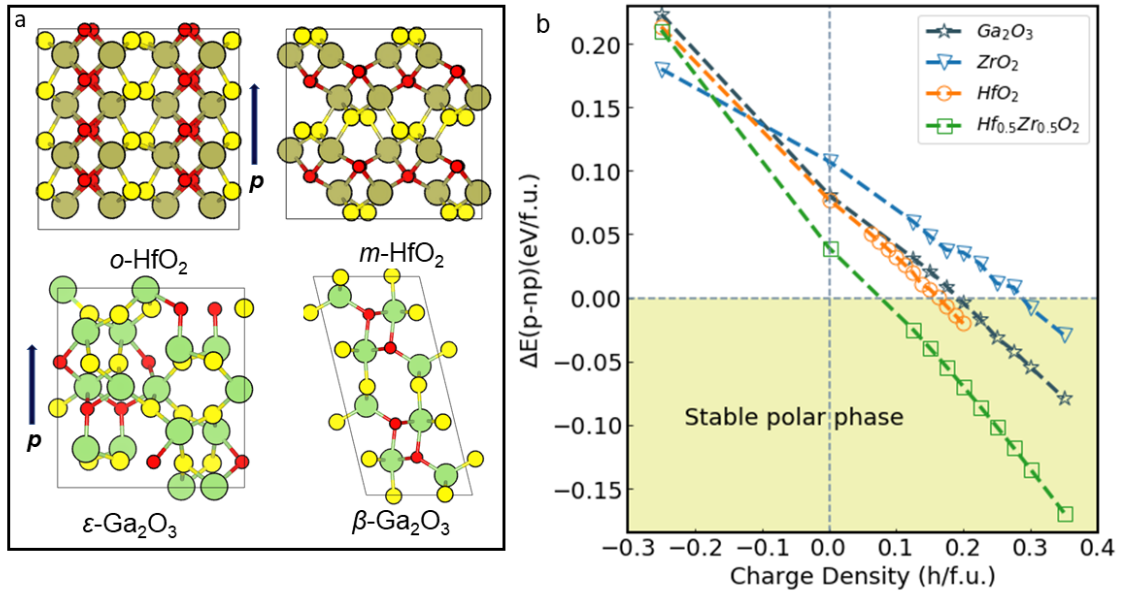

**Fig. 1** (a) Crystal structures of $o$-$HfO_2$, $m$-$HfO_2$, $\varepsilon$-$Ga_2O_3$, and $\beta$-$Ga_2O_3$. Here filled circles in olive green and light green are Hf and Ga atoms, and circles in yellow and red are O3 and O4 atoms, respectively. (b) Energy difference between polar and non-polar phases for different binary metal oxides as a function of charge carrier density.

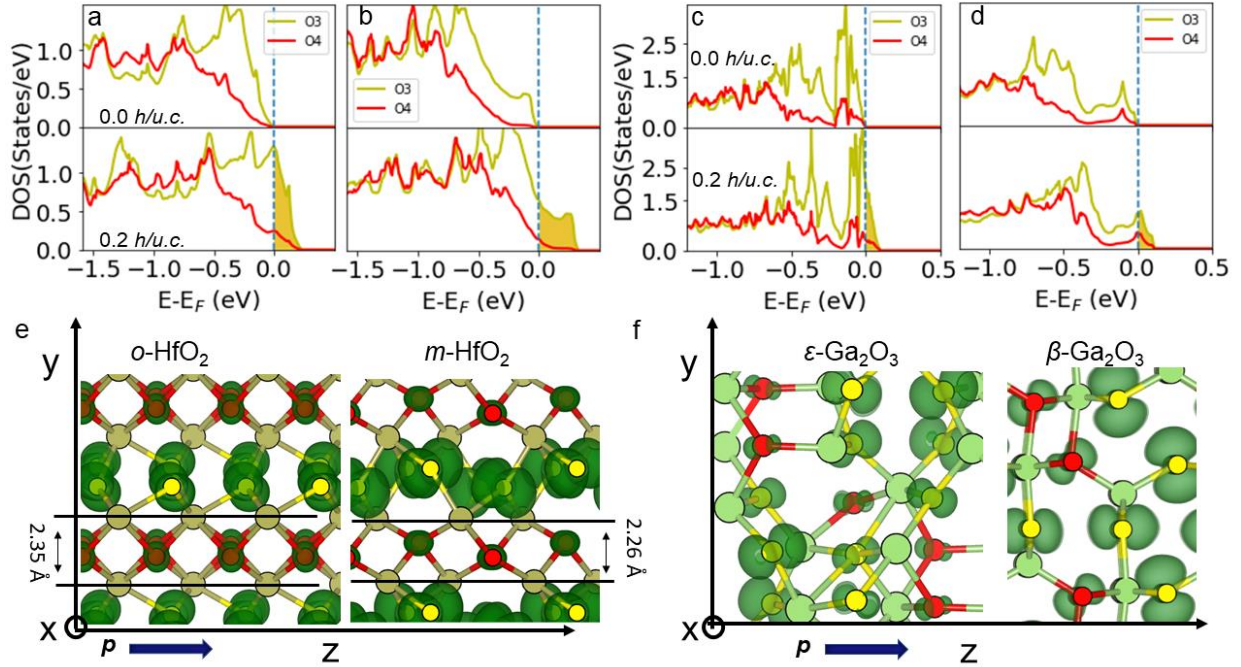

**Fig. 2** (a-d) Projected density of states (DOS) on O3 and O4 atoms in charge neutral and 0.2 h/f.u. hole-doped state for $o$-HfO$_2$ (a), $m$-HfO$_2$ (b), $\varepsilon$-Ga$_2$O$_3$ (c), and $\beta$-Ga$_2$O$_3$ (d). (e-f) Real-space distribution of the hole density (0.2 h/f.u.) in $o$-, $m$-HfO$_2$ (e) and $\varepsilon$-, $\beta$-Ga$_2$O$_3$ (f). The isosurface of the hole density is at 10 % of its maximum.

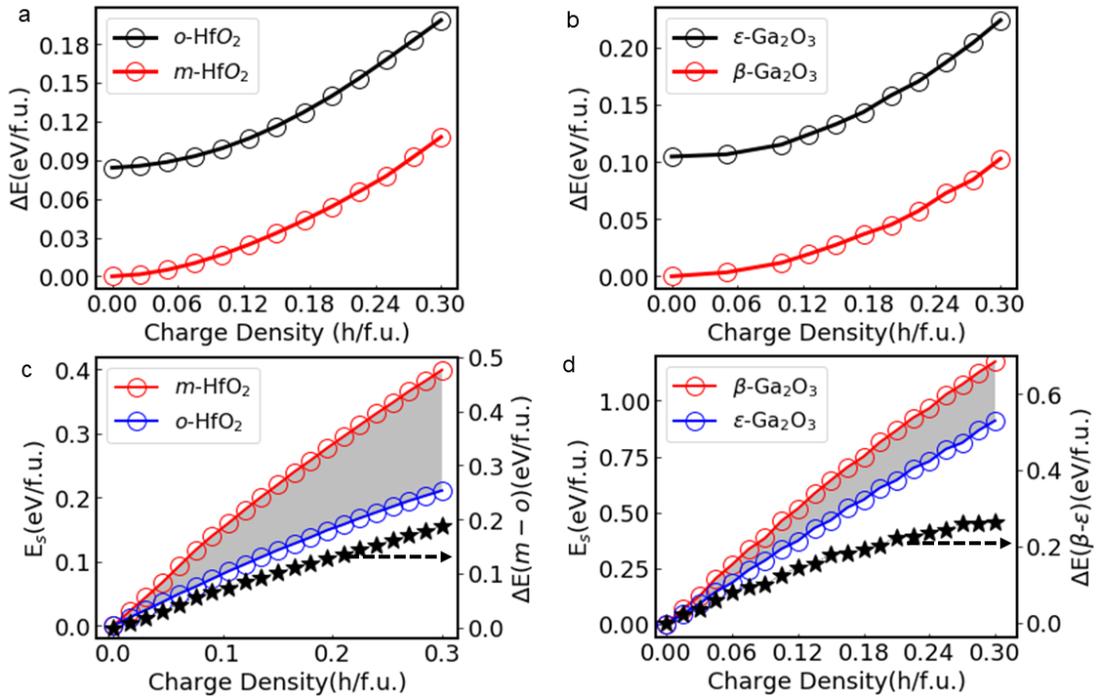

**Fig. 3** (a-b) The deformation energy for $o$-$HfO_2$, $m$-$HfO_2$ (a), and $\varepsilon$-$Ga_2O_3$, $\beta$-$Ga_2O_3$ (b) as a function of hole density. (c-d) The electrostatic energy variation and electrostatic energy difference between nonpolar and polar phase in $HfO_2$ and $Ga_2O_3$ as a function of the hole density.

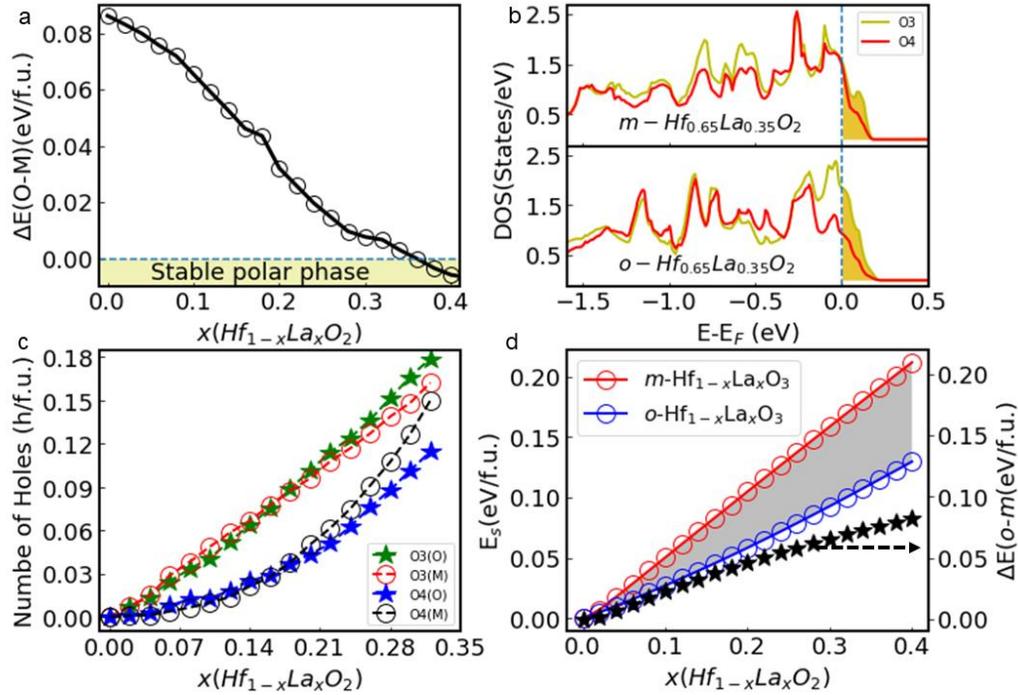

**Fig. 4** (a) Energy difference between ferroelectric $o$- and monoclinic $m$-$Hf_{1-x}La_xO_2$ as a function of La concentration $x$. (b) The density of states (DOS) of O3 and O4 in $o$- and $m$-$Hf_{0.65}La_{0.35}O_2$. (c) The number of holes at O3 and O4 sites in $o$- and $m$-$Hf_{1-x}La_xO_2$. (d) The electrostatic energy of $o$- and $m$-$Hf_{1-x}La_xO_2$ and the electrostatic energy difference as a function of La concentration.

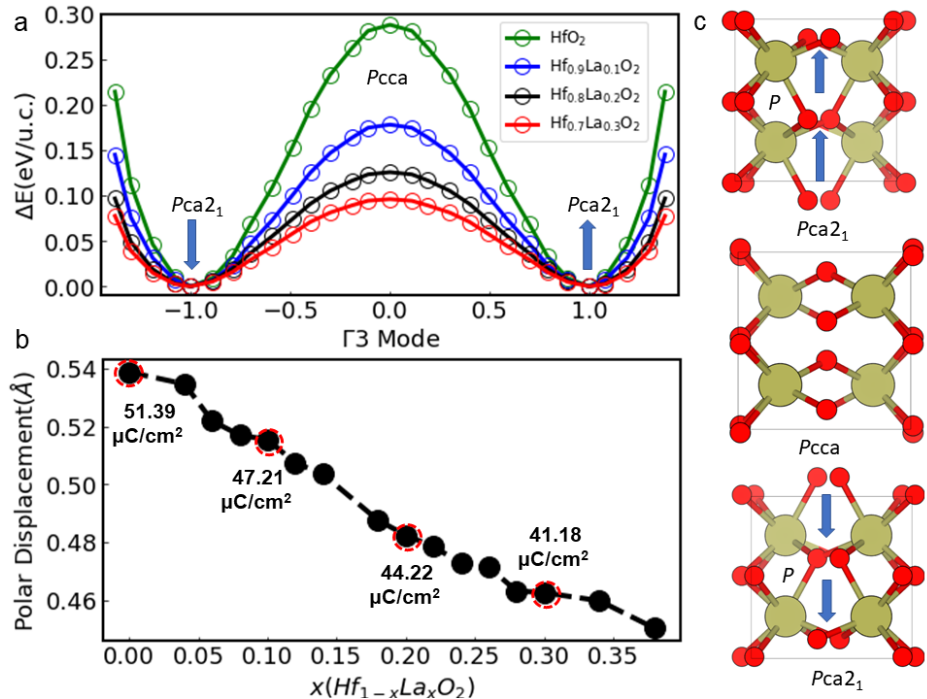

**Fig. 5** (a) Energy of the ferroelectric double-well for $HfO_2$, $Hf_{0.9}La_{0.1}O_2$, $Hf_{0.8}La_{0.2}O_2$, and $Hf_{0.7}La_{0.3}O_2$. (b) Polar displacement in $HfO_2$ as a function of La doping concentration. (c) Schematic of the ferroelectric switching transition from $Pca2_1$-phase of $HfO_2$ with the polarization pointing up (top) to the state with polarization pointing down (bottom) through a centrosymmetric $Pcca$ phase (middle).

**Acknowledgements**

Tengfei Cao thanks Prof. Angelo Bongiorno at the CUNY College of Staten Island for helpful discussions. The work at Washington University was supported by the National Science Foundation (NSF) through awards DMR-1806147, DMR-1931610 and DMR-2145797. The work at University of Nebraska-Lincoln was partly supported by the NSF EPSCoR RII Track-1 program (Award OIA-2044049). Computations were performed at the Extreme Science and Discovery Environment (XSEDE), supported by NSF ACI-1548562, and at the University of Nebraska Holland Computing Center.